\documentclass[twocolumn,showpacs,preprintnumbers,amsmath,amssymb,superscriptaddress,prl]{revtex4}
\usepackage{graphicx}
\usepackage{dcolumn}
\usepackage{bm}

\begin{document}

\title{Precision measurement of the quasi--free $pn{\to}d\phi$ reaction close to
threshold}

\author{Y.\,Maeda}
\affiliation{Institut f\"ur Kernphysik, Forschungszentrum
J\"ulich, 52425 J\"ulich, Germany}%
\affiliation{Research Center for Nuclear Physics, Osaka
University, Ibaraki, Osaka 567-0047, Japan}
\author{M.\,Hartmann}\email[E-mail: ]{M.Hartmann@fz-juelich.de}%
\affiliation{Institut f\"ur Kernphysik,
Forschungszentrum J\"ulich, 52425 J\"ulich, Germany}
\author{I.\,Keshelashvili}
\affiliation{Institut f\"ur Kernphysik, Forschungszentrum
J\"ulich, 52425 J\"ulich, Germany}
\affiliation{High Energy Physics Institute, Tbilisi State
University, 0186 Tbilisi, Georgia}
\author{S.\,Barsov}
\affiliation{High Energy Physics Department, Petersburg
Nuclear Physics Institute, 188350 Gatchina, Russia}
\author{M.\,B\"uscher}
\affiliation{Institut f\"ur Kernphysik, Forschungszentrum
J\"ulich, 52425 J\"ulich, Germany}
\author{A.\,Dzyuba}
\affiliation{Institut f\"ur Kernphysik, Forschungszentrum
J\"ulich, 52425 J\"ulich, Germany}
\affiliation{High Energy Physics Department, Petersburg
Nuclear Physics Institute, 188350 Gatchina, Russia}
\author{S.\,Dymov}
\affiliation{Laboratory of Nuclear Problems, Joint Institute
for Nuclear Research, 141980 Dubna, Russia}
\author{V.\,Hejny}
\affiliation{Institut f\"ur Kernphysik, Forschungszentrum
J\"ulich, 52425 J\"ulich, Germany}
\author{A.\,Kacharava}
\affiliation{High Energy Physics Institute, Tbilisi State
University, 0186 Tbilisi, Georgia}
\affiliation{Physikalisches Institut II, Universit\"at
  Erlangen--N\"urnberg, 91058 Erlangen, Germany}
\author{V.\,Kleber}
\affiliation{Physikalisches Institut, Universit\"at
Bonn, 53115 Bonn, Germany}
\author{H.R.\,Koch}
\affiliation{Institut f\"ur Kernphysik, Forschungszentrum
J\"ulich, 52425 J\"ulich, Germany}
\author{V.\,Koptev}
\affiliation{High Energy Physics Department, Petersburg
Nuclear Physics Institute, 188350 Gatchina, Russia}
\author{P.\,Kulessa}
\affiliation{Institut f\"ur Kernphysik, Forschungszentrum
J\"ulich, 52425 J\"ulich, Germany}
\affiliation{H.~Niewodnicza\'{n}ski Institute of Nuclear
Physics PAN, 31342 Krak\'{o}w, Poland}
\author{T.\,Mersmann}
\affiliation{Institut f\"ur Kernphysik, Universit\"at
M\"unster, 48149 M\"unster, Germany}
\author{S.\,Mikirtytchiants}
\affiliation{High Energy Physics Department, Petersburg
Nuclear Physics Institute, 188350 Gatchina, Russia}
\author{A.\,Mussgiller}
\affiliation{Institut f\"ur Kernphysik, Forschungszentrum
  J\"ulich, 52425 J\"ulich, Germany}
\affiliation{Physikalisches Institut II, Universit\"at
  Erlangen--N\"urnberg, 91058 Erlangen, Germany}
\author{M.\,Nekipelov}
\affiliation{Institut f\"ur Kernphysik, Forschungszentrum
  J\"ulich, 52425 J\"ulich, Germany}
\author{H.\,Ohm}
\affiliation{Institut f\"ur Kernphysik, Forschungszentrum
  J\"ulich, 52425 J\"ulich, Germany}
\author{R.\,Schleichert}
\affiliation{Institut f\"ur Kernphysik, Forschungszentrum
  J\"ulich, 52425 J\"ulich, Germany}
\author{H.J.\,Stein}
\affiliation{Institut f\"ur Kernphysik, Forschungszentrum
  J\"ulich, 52425 J\"ulich, Germany}
\author{H.\,Str\"oher}
\affiliation{Institut f\"ur Kernphysik, Forschungszentrum
  J\"ulich, 52425 J\"ulich, Germany}
\author{Yu.\,Valdau}
\affiliation{High Energy Physics Department, Petersburg Nuclear
Physics Institute, 188350 Gatchina, Russia}
\author{K.--H.\,Watzlawik}
\affiliation{Institut f\"ur Kernphysik, Forschungszentrum
  J\"ulich, 52425 J\"ulich, Germany}
\author{C.\,Wilkin}
\affiliation{Physics and Astronomy Department, UCL,
Gower Street, London WC1E 6BT, UK}
\author{P.\,W\"ustner}
 \affiliation{Zentralinstitut f\"ur Elektronik,
Forschungszentrum J\"ulich, 52425 J\"ulich, Germany}
\date{\today}

\begin{abstract}
The quasi--free $pn{\to}d{\phi}$ reaction has been studied at the
Cooler Synchrotron COSY--J\"ulich, using the internal proton beam
incident on a deuterium cluster--jet target and detecting a fast
deuteron in coincidence with the $K^+K^-$ decay of the
$\phi$--meson. The energy dependence of the total and differential
cross sections are extracted for excess energies up to 80~MeV by
determining the Fermi momentum of the target neutron on an
event--by--event basis. Though these cross sections are consistent
with $s$--wave production, the kaon angular distributions show the
presence of $p$ waves at quite low energy. Production on the
neutron is found to be stronger than on the proton but not by as
much as for the $\eta$--meson.
\end{abstract}

\pacs{25.40.Ve, 13.75.Cs, 14.40.Cs}%
\maketitle
Meson production provides access to the internal structure of
baryons and the dynamics of hadronic reactions and thus is an
important exploration field for non--perturbative QCD. In
proton--proton collisions, meson production has been extensively
studied and data are now available on the production of most
members of the fundamental pseudoscalar and vector nonets near
their respective threshold~\cite{Hanhart}, including the
$\pi(140)$, $\eta(547)$, and $\eta^{\prime}(958)$, as well as the
$\rho(770)$, $\omega(782)$, and $\phi(1020)$. The $\phi$--meson is
of particular interest because of its comparatively large mass and
its dominant $s\bar{s}$ quark structure.  However, in order to
study all facets of meson production dynamics, it is necessary to
investigate the isospin dependence by precision measurements in
both $pn$ as well as $pp$ collisions. In the case of the
$\eta$--meson, such experiments have revealed that the $pn$
production cross section is over six times larger than that for
$pp$~\cite{Stina}. Analogous $\phi$ data are important for
nucleon--nucleon production models and also serve as crucial input
in the interpretation of nucleon-nucleus, and nucleus-nucleus
results, where \emph{in--medium} effects are
anticipated~\cite{medium}.

In the absence of a free neutron target or a quality neutron beam,
quasi--free production on deuterium $pd\,{\to}\,d X \,p_{\rm sp}$
has often been substituted. Here the reaction is assumed to have
taken place on the neutron bound in the deuteron and $p_{\rm sp}$
is a slow ``spectator'' proton that does not take an active part
in the reaction and whose momentum reflects the Fermi motion of
the particle before the production. In order to show that the
reaction involved only the neutron, the spectator must be
identified and the precise determination of the c.m.\ energy
requires that the $p_{\rm sp}$ momentum is well measured.
Spectators emerging from an ultra--thin target with a few MeV can
be studied directly at a storage ring using solid--state counters,
as has been done for $pd\,{\to}\,d \pi^0\,p_{\rm sp}$~\cite{Tord}
and $pd\,{\to}\,d\,\omega \,p_{\rm sp}$~\cite{Barsov}. The
alternative approach is to identify the produced meson $X$ through
its decay products and then reconstruct the spectator momentum
using kinematics. This method has been successfully employed for
the $pd\,{\to}\,d \eta\,p_{\rm sp}$ and $pd\,{\to}\,pn
\eta\,p_{\rm sp}$ reactions, where the $\eta$ was identified
through its $2\gamma$ decay branch~\cite{Stina}. We have studied
for the first time quasi--free $pn\to d\phi$ production through
the indirect method of measuring the spectator momentum using the
$K^+K^-$ decay of the $\phi$ in coincidence with a fast deuteron.

The experiment was performed with a $2.65\,$GeV proton beam at an
internal target station of the Cooler Synchrotron COSY, employing
the magnetic spectrometer ANKE~\cite{ANKE} to identify and measure
the reaction. ANKE has detection systems placed to the right and
left of the emerging beam to register slow positively and
negatively charged ejectiles, with fast positively charged
particles being measured in the forward system. The deuterium
cluster--jet target~\cite{GasTarget} provided areal densities of
\mbox{$\sim 3.4\!\times\!10^{14}\,\textrm{cm}^{-2}$} which,
combined with a typical proton beam intensity of \mbox{$\sim 6.2
\times 10^{16}\,\textrm{s}^{-1}$}, gave an integrated luminosity
of 23\,pb$^{-1}$ over the 300 hours of data taking.

The $pd\,{\to}\,d \phi \,p_{\rm sp}$ reaction was studied in a
manner analogous to that successfully employed for the
$pp\,\to\,pp\phi$ reaction at COSY~\cite{ANKEpp}, using the
$\phi\to K^+K^-$ decay. Charged kaon pairs were detected in
coincidence with a forward--going deuteron, requiring that the
overall missing mass in the reaction was consistent with that of
the non--observed slow spectator proton $p_{\rm sp}$. As a first
step, positive kaons are selected through a procedure described in
detail in Ref.~\cite{ANKEKAON}, using the time of flight (TOF)
between START and STOP scintillation counters of a dedicated $K^+$
detection system. In the second stage, both the coincident $K^{-}$
and forward--going deuteron are identified from the
time--of--flight differences between the STOP counters in the
negative and forward detector systems with respect to the STOP
counter in the positive system that was hit by the $K^+$. These
two TOF selections, as well as that for the $K^+$, were carried
out within $\pm\,3\,\sigma$ bands.

\begin{figure}[ht]
  \vspace*{+1mm}
  \centering
  \includegraphics[clip,width=0.23\textwidth]{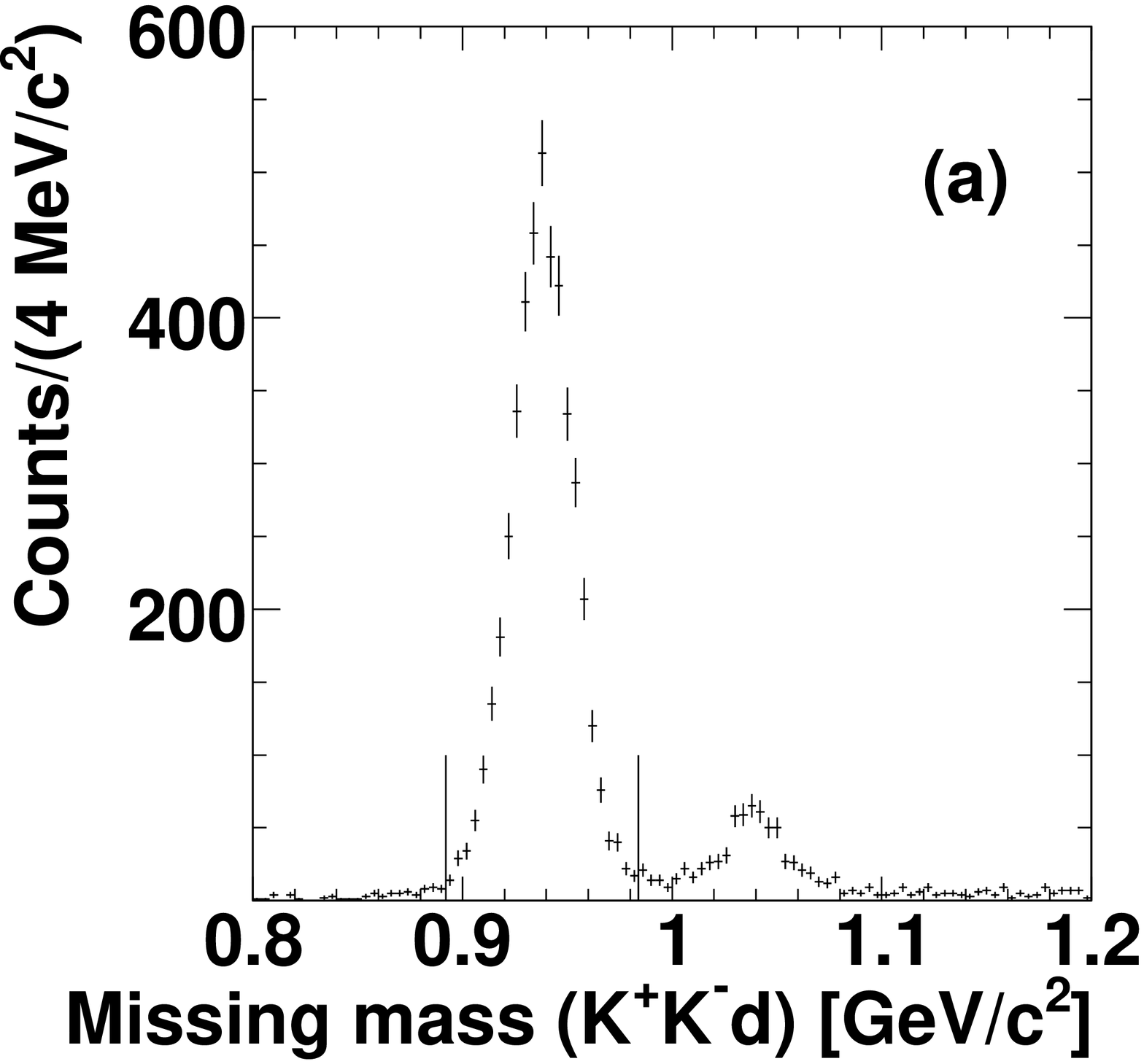}
  \includegraphics[clip,width=0.23\textwidth]{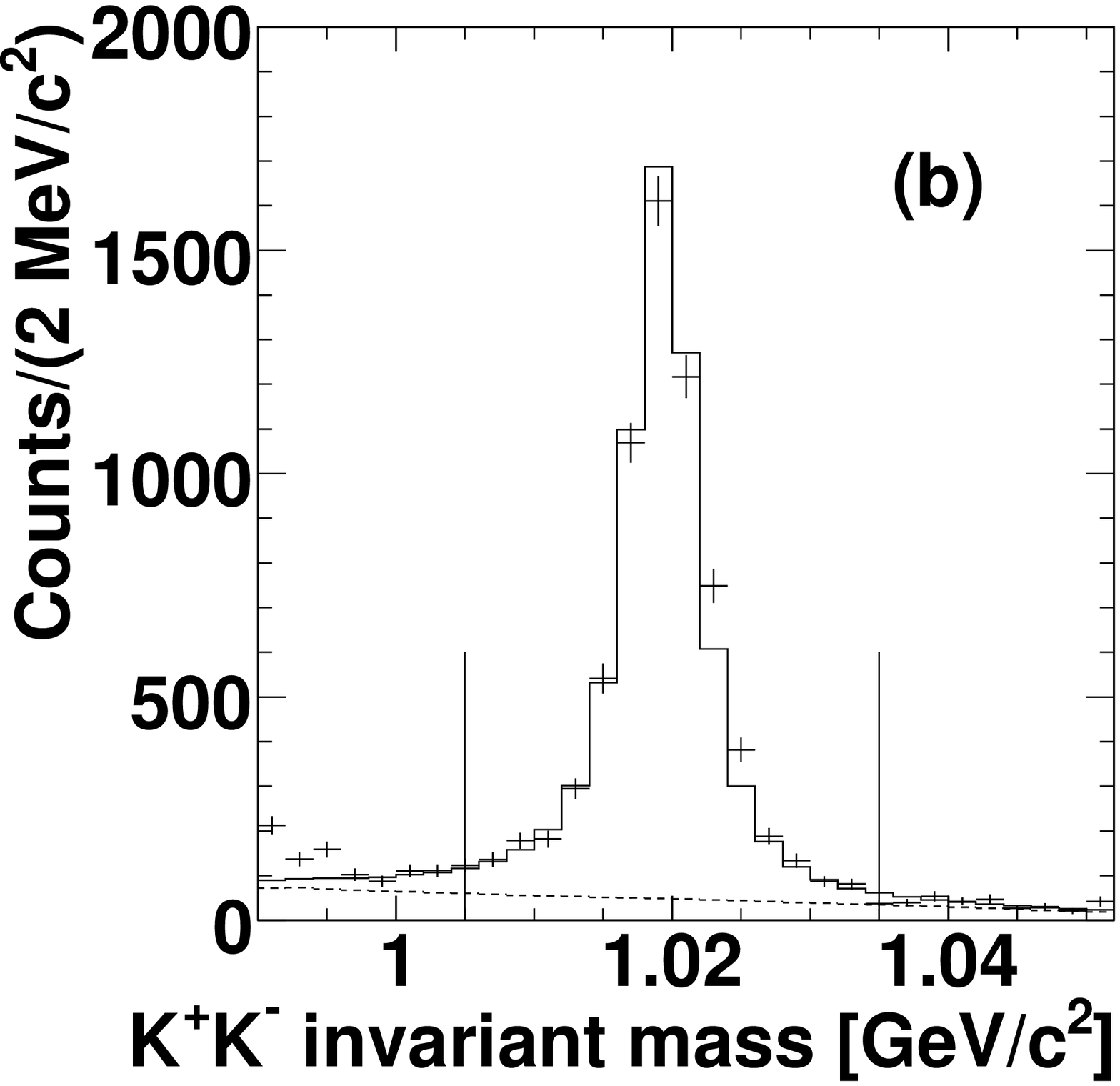}
  \includegraphics[clip,width=0.23\textwidth]{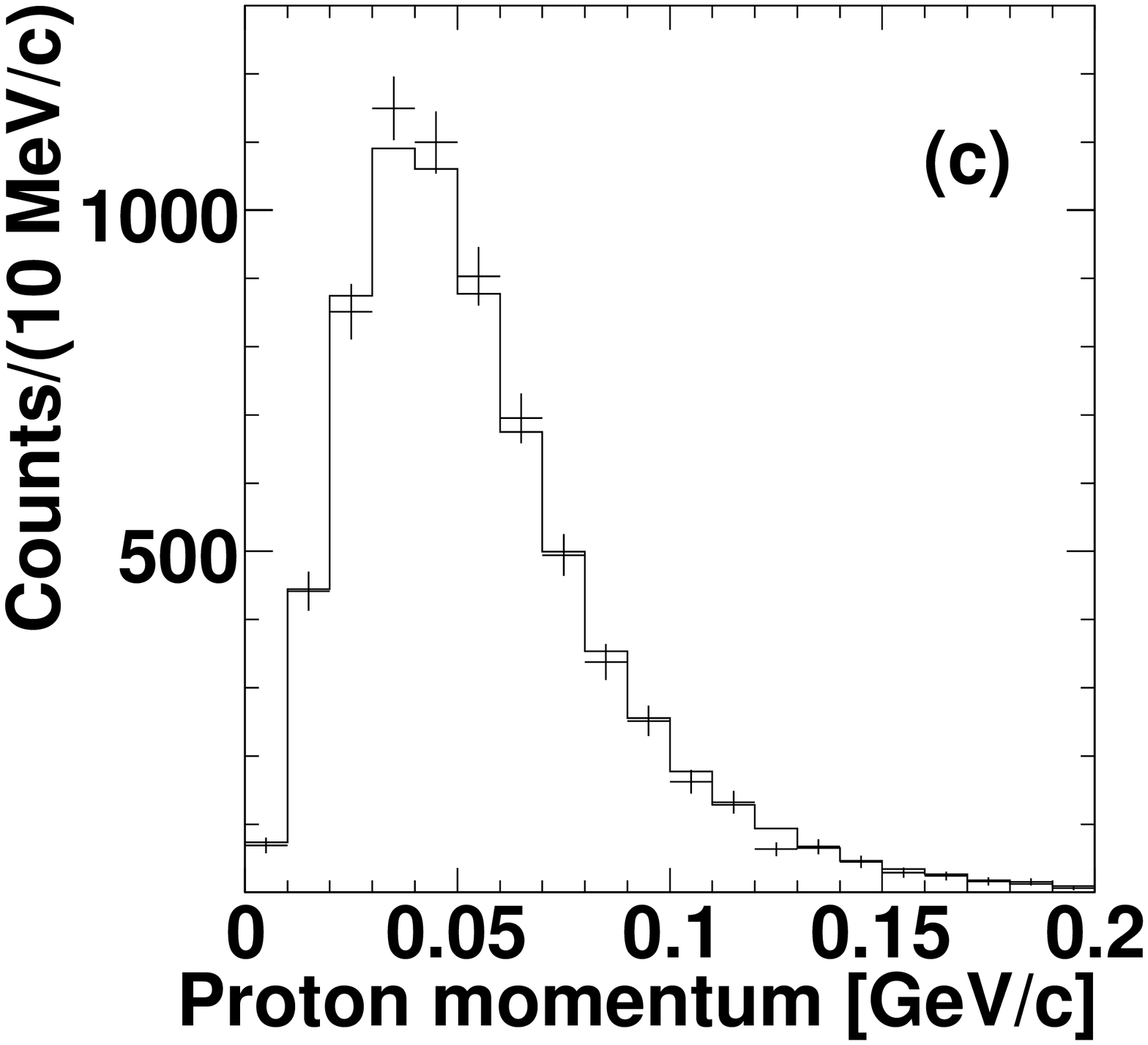}
  \includegraphics[clip,width=0.23\textwidth]{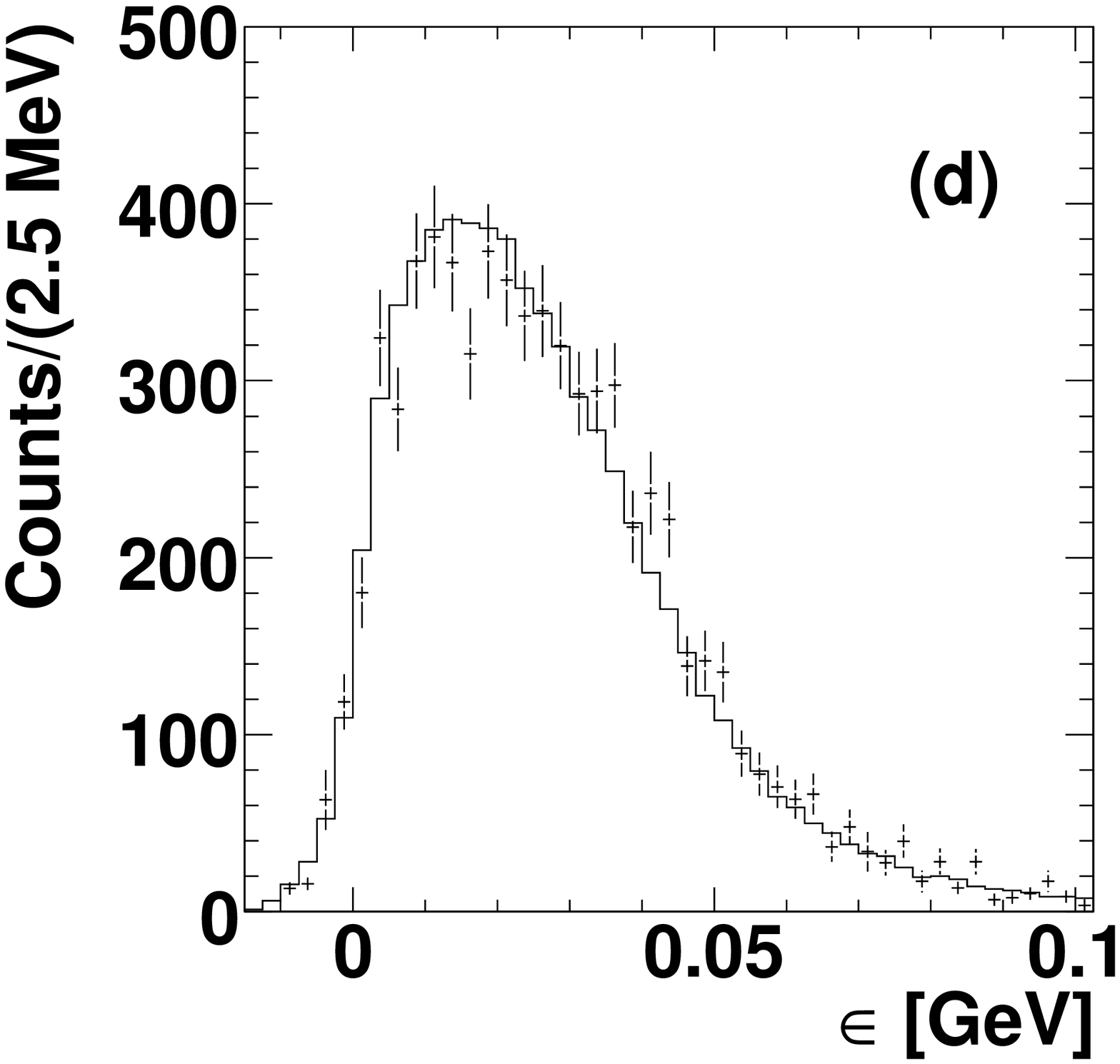}
  \vspace*{-2mm}
  \caption{
a) Missing--mass distribution of the $pd\,{\to}\,d K^{+}K^{-}\,X$
events with lines indicating the proton selection range. b)
$K^{+}K^{-}$ invariant--mass distribution with lines showing the
mass range of selected $\phi$--meson events. The dashed curve
shows the four--body phase--space fit to the non--resonant
production while the solid histogram is the sum of this and the
$\phi$--meson contribution. c) Momentum distribution of the
unobserved proton compared with a Monte--Carlo simulation based on
the spectator model. d)  Distribution in c.m.\ excess energy compared
with the same simulation. }\label{IM} \vspace*{-2mm}
\end{figure}

Fig.~\ref{IM}a, which shows the missing mass spectrum assuming
that the detected particles are indeed $K^+$, $K^-$ and deuteron,
demonstrates a clear peak at the mass of the proton. The secondary
peak around 1.02~GeV/c$^2$ arises from $p\pi^+\pi^-$ events, where
a $\pi^+$ was misidentified as a $K^+$. This background is well
separated from the spectator peak over the whole kinematic region.
The residue from misidentified particles inside the proton gate
(of $\pm\,3\,\sigma$) is $3.1\,\%$ and such events generally fail
the later criteria of the analysis. In total, about 4500
coincidences were retained as $dK^+K^- p_{\rm sp}$ events for
further study.

The $K^+K^-$ invariant--mass spectrum for the 4500 events is shown
in Fig.~\ref{IM}b. The distribution is dominated by the
$\phi$--meson peak, which sits on a slowly varying physical
background from direct $K^{+}K^{-}$ production. This has been
estimated by a four--body phase--space simulation which, together
with the $\phi$ contribution, is fitted to the overall spectrum.
The shape of the resonant contribution is reproduced by the
natural width of $\phi$--meson with an experimental mass resolution
$\sigma=1~\textrm{MeV/c}^2$, which is consistent with the
momentum resolution of the ANKE detector system. The direct
$K^+K^-$ contribution, which is less than $8\%$ in the $\phi$ mass
region $1.020\pm0.015~\textrm{GeV/c}^2$, could be easily
subtracted.

The momentum distribution of the unobserved proton for events in
the $\phi$ peak is shown in Fig.~\ref{IM}c. As expected for a
spectator proton, this spectrum is peaked at very low values and
there are few events with momenta above about 150~MeV/c. To
confirm the spectator hypothesis, a Monte Carlo simulation has
been performed where the Fermi momentum in the target deuteron has
been derived from the Bonn potential~\cite{BONN}. The energy
dependence of the $pn\to d\phi$ cross section is assumed to follow
phase space, which is consistent with the results to be shown
later. After including the detector response, the simulation fits
very well the shape of the data for momenta up to at least
150~MeV/c, a region where the model dependence of the deuteron
wave function is negligible compared with our statistical
uncertainty. The spectator distribution could be obtained with
even greater precision than that for the corresponding $pd\,\to\,
d\eta\, p_{\rm sp}$ reaction~\cite{Stina}.

Due to the Fermi motion of the neutron in the deuteron, the c.m.\
excess energy $\epsilon=\sqrt{s}-(m_{d}+m_{\phi})$ is spread over
a range of values even for a fixed beam energy. Since we have
completely determined the kinematics for each of the $pd\,{\to}\,d
\phi \,p_{\rm sp}$ events, the value of $\epsilon$ could be
calculated on an event--by--event basis, and the resulting
distribution is shown in Fig.~\ref{IM}d. This is also well
described by the simulation, which shows that $\epsilon$ can be
reconstructed with an average precision of $\sigma_{\epsilon}
=2\,$MeV, and which can be used in the extraction of cross
sections for $\epsilon < 80\,$MeV.

The target density was determined by measuring the frequency shift
of the stored proton beam as it lost energy due to its repeated
passages through the target~\cite{Zapfe}. Combined with
measurements of the beam current this yielded the value of the
luminosity $L$ with a precision of about $\pm 6\,$\%. This was
checked through the simultaneous measurement of $pd$ elastic and
quasi--elastic scattering, where a fast proton was registered in
the polar angular range $5.0^{\circ}<\vartheta < 8.5^{\circ}$ in
the forward detectors. The luminosity was then obtained from
estimates~\cite{Uzikov} of the corresponding differential cross
sections within the Glauber formalism~\cite{FRANCO}. This
technique has been used successfully at other
energies~\cite{Komarov}. Though the two methods give consistent
results to within 3\%, the error in the $pd$ technique is about
10\%, due mainly to the use of the theoretical model and
uncertainty in the acceptance correction.

\begin{figure}[t]
\vspace*{+0.9mm}
  \centering
  \includegraphics[clip,width=0.46\textwidth]{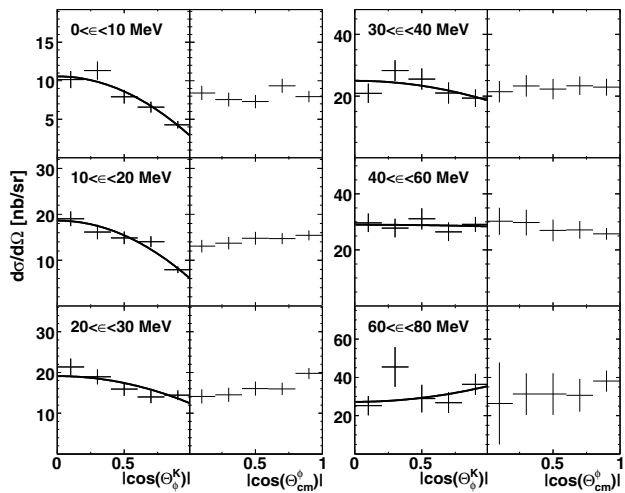}
  \vspace*{-2mm}
  \caption{
Differential cross sections for $pn\to d\phi$ for different ranges
of excess energy, with the left panels showing the dependence on
the angle of the $K^+$ from the $\phi$ decay in the $\phi$ rest
frame, and the right panels the dependence on the polar angle of
the $\phi$ in the overall c.m.\ system. Vertical error bars
indicate purely statistical uncertainties whereas the horizontal
ranges reflect the bin width.
  }
\label{DXS} \vspace*{-1mm}
\end{figure}

In order to evaluate differential cross sections, the geometrical
acceptance, resolution, detector efficiency and kaon decay
probability were taken into account in a Monte Carlo simulation,
using the GEANT4 package~\cite{GEANT4}. For a given excess energy,
the distributions in all variables were consistent with phase
space except for that of $\Theta_{\phi}^{K}$, which is the polar
angle of the $K^+$ from the $\phi$ decay with respect to the beam
direction, in the $\phi$ rest frame. At threshold the only allowed
$pn\to d\phi$ transition arises from an initial $^1P_1$ state. The
unique production amplitude is therefore of the form
$M=\bm{p}\cdot(\bm{\varepsilon}_{d}^{\dagger}\times
\bm{\varepsilon}_{\phi}^{\dagger})\,\Phi_{pn}$, where $\bm{p}$ is
the beam momentum, $\bm{\varepsilon}_{d}$ and
$\bm{\varepsilon}_{\phi}$ are the polarization vectors of the
deuteron and $\phi$ respectively, and $\Phi_{pn}$ represents the
spin--0 initial $pn$ state. From the structure of the matrix
element, it follows that the $\phi$--meson is aligned
transversally to the beam so that, following its decay, the kaons
cannot be produced along the beam direction and a
$\sin^{2}\Theta^{K}_{\phi}$ behavior is to be expected.

To allow for the possibility of higher partial waves, the
distribution was parameterized in the most general allowed form:
$\textrm{d}\sigma/\textrm{d}\Omega_{\phi}^{K} =
3(a\sin^{2}\Theta^{K}_{\phi} + 2 b
\cos^{2}\Theta^{K}_{\phi})/8\pi$, normalized such that the total
cross section $\sigma=a+b$. This form was handled iteratively in
the simulation to get the best values of the parameters $a$ and
$b$ and of the ANKE acceptance. For large $\epsilon$ the
acceptance in the backward c.m.\ hemispheres are somewhat higher
than in the forward, but all distributions are completely
consistent with them being symmetric in the c.m.\ system. The
results for different excess energy bins in Fig.~\ref{DXS} are
therefore shown as functions of the magnitudes of the cosines of
$\Theta^{K}_{\phi}$ and the polar angle $\Theta^{\phi}_{cm}$ of
the $\phi$, for which the resolutions are estimated to be $0.024$
and typically $0.02$--$0.04$ respectively. The dominance of the
$\sin^{2}\Theta^{K}_{\phi}$ term is very clear at the lower
energies and all the data are well represented by $b/a\approx
(0.012\pm 0.001)\,(\epsilon/\textrm{MeV})$. Given the ambiguity
associated with the nine possible $p$--wave amplitudes, this ratio
represents the minimum fraction of higher partial waves and
indicates that this is significant for the larger $\epsilon$.
Despite this, the angular distribution of $\phi$ production in the
overall c.m.\ system is consistent with isotropy for all
$\epsilon$. Note that the production of $p$--wave $\phi$ from an
initial $^3S_1$ state would also be flat in
$\cos\Theta_{cm}^{\phi}$.

\begin{figure}[t]
  \centering
  \vspace*{+1.2mm}
  \includegraphics[clip,width=0.5\textwidth]{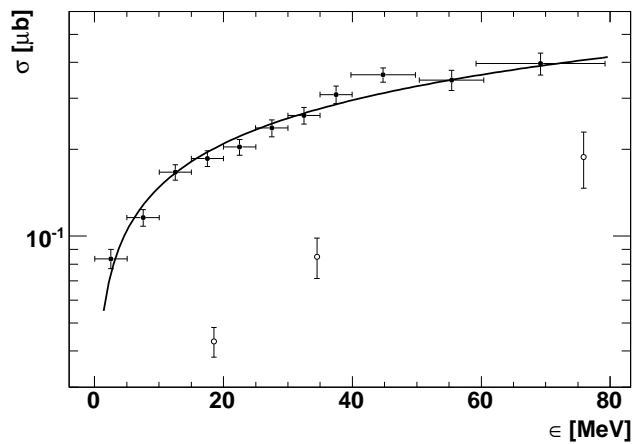}
  \vspace*{-2mm}
  \caption{
Total cross section for the quasi--free $pn\to d\phi$ reaction as
a function of the excess energy (filled circles). In addition to
the statistical error bars shown, there are overall systematic
uncertainties of $\pm 10$\%. The curve represents a phase--space
$\sqrt{\epsilon}$ behavior. For comparison, we also show as open
circles the values obtained from $pp\to pp\phi$~\cite{ANKEpp}.
  }
\label{Diff19MeV}
\vspace*{-1mm}
\end{figure}

The values of the $a$ and $b$ coefficients lead directly to the
total cross section for $\phi$ production shown in
Fig.~\ref{Diff19MeV}, while numerical values will be found in the
HEP database~\cite{Durham}. In addition to the point--to--point
statistical errors, there is an overall systematic uncertainty of
$\pm 10$\% coming from luminosity (6\%), stability of the
data--taking efficiency (4\%), background (3\%), and MWPC
efficiency corrections for kaon detection (4\%). The results have
not been corrected for the reduction in the incident flux due to
shadowing by the proton in the deuteron target, which would
increase the cross sections by about 4\%~\cite{Chiavassa}. Values
of the $pp\to pp\phi$ total cross sections available in our energy
range are also shown~\cite{ANKEpp}.

Two--body phase space increases like $\sqrt{\epsilon}$ and this is
distorted by less than 4\% when the width of the $\phi$--meson is
taken into account. As shown in Fig.~\ref{Diff19MeV}, we find that
$\sigma(pn\to d\phi)=(48\pm 1)\,
\sqrt{(\epsilon/\textrm{MeV})}~\textrm{nb}$, despite the decay
angular distributions showing significant $p$--wave effects at
higher $\epsilon$. The values are much higher than those of $pp\to
pp\phi$~\cite{ANKEpp,DISTO}, but this is due in part to there
being a three-- rather than a two--body final state. However, very
near threshold, isoscalar $S$--wave $\phi pn$ production can be
estimated from our $d\phi$ data using final--state--interaction
theory~\cite{FW}, a technique that has been tested for $\eta$
production~\cite{Stina}. This approach yields $\sigma(pn\to
pn\phi)/\sigma(pp\to pp\phi)\approx 2.3\pm 0.5$, which is only
about a third as big as the ratio for $\eta$
production~\cite{Stina}.

The ratio $R_{\phi/\omega}$ of the production of the light
isoscalar vector mesons $\phi$ and $\omega$ in various nuclear
reactions involving non--strange particles provides valuable tests
of the Okubo--Zweig--Iizuka rule~\cite{OZI}. This rule suggests
that, due to small deviations from ideal mixing of these mesons,
one should have $R_{\phi/\omega}\approx R_{\textrm OZI} =
4.2\times 10^{-3}$~\cite{Lipkin} under similar kinematic
conditions. Significant enhancements of this ratio have, however,
been reported in the literature and for proton--proton collisions
we recently obtained $\sigma(pp\to pp\phi)/\sigma(pp\to pp\omega)
= (3.3\pm 0.6)\times 10^{-2} \approx 8\times R_{\rm
OZI}$~\cite{ANKEpp}. There is a measurement of $\omega$ production
in proton--neutron collisions at
$57^{+21}_{-15}$~MeV~\cite{Barsov} and, comparing this with our
data, we find that at this energy  $\sigma(pn\to
d\phi)/\sigma(pn\to d\omega) = (4.0\pm 1.9)\times 10^{-2} \approx
9\times R_{\rm OZI}$. Though similar to the $pp$ result, the error
bar is large.

In near--threshold production reactions, the relevant degrees of
freedom seem to be mesons and baryons rather than quarks and
gluons, and the predictions for $\phi$ production in $pn\to d\phi$
are very sensitive to meson exchange and nucleonic
currents~\cite{Nakayama,Kondratyuk,Kampfer}. Nevertheless, all
three calculations yield broadly similar values for the $pn\to
d\phi$ total cross section, being in the $0.1$--$0.5\,\mu$b range
at $\epsilon\approx 50\,$MeV compared with the $\approx
0.34\,\mu$b shown by our data in Fig.~\ref{Diff19MeV}. However,
whereas one calculation suggests that the $\phi$ production is
maximal in the forward direction~\cite{Kampfer}, another predicts
it to be much flatter~\cite{Nakayama}. Our data in Fig.~\ref{DXS}
are consistent with isotropy. Although no calculations appear to
exist in the literature for the polarization of the $\phi$--meson
in the $pn\to d\phi$ reaction, estimates have been made for that
of the deuteron and of the initial $pn$
spin--correlation~\cite{Kampfer}. Both of these observables are
sensitive to effects from $p$--wave $\phi$ production but neither
shows as big effects with energy as we have seen from the $\phi$
alignment in Fig.~\ref{DXS}.

In summary, we have presented the first measurements of $\phi$
production in $pn$ collisions that will provide important
constraints on the theoretical modeling of such processes. The
$\phi$ polarization, as measured through its $K^+K^-$ decay, shows
the early onset of $p$--waves which are not apparent in the energy
variation of the total cross section. This behavior can also not
be seen in the c.m.\ angular distributions and this suggests that
$p$--waves might be more important than previously thought in
other near--threshold meson production. The production is stronger
in $pn$ collisions than in $pp$, though the factor is not as large
as for $\eta$ production.

These data are, of course, valuable in the interpretation of
$\phi$--meson production in the collision of heavy ions. Further
testing of the OZI rule in $pn$ collisions will have to await
better data on $\omega$ production. This should be possible in the
future at the WASA at COSY facility~\cite{WASA}, where photons
from the $\omega\to\pi^0\gamma$ decay can be detected in
coincidence with fast deuterons. This decay mode would also allow
the polarization of the $\omega$ to be analyzed to see if this
also shows a rapid onset of $p$ waves.

Support from J.\,Haidenbauer, C.\,Hanhart, U.--G.\ Mei\ss{}ner,
A.\,Sibirtsev, Yu.\,Uzikov, K.\,Nakayama, and other members of the
ANKE Collaboration, as well as the COSY machine crew, are
gratefully acknowledged. This work has been partially financed by
the BMBF, DFG, Russian Academy of Sciences, and COSY FFE.


\end{document}